\documentclass[%
 reprint,
 amsmath,amssymb,
 aps,
]{revtex4-2}

\usepackage{graphicx}
\usepackage{dcolumn}
\usepackage{bm}

\begin{document}

\preprint{APS/123-QED}

\title{Beyond the maximum drag reduction asymptote: \\
the pseudo-laminar state}

\author{Anselmo Pereira}
\thanks{anselmo.soeiro\_pereira@mines-paristech.fr}
\affiliation{PSL Research University, MINES ParisTech, Centre for material forming (CEMEF), CNRS UMR 7635, CS 10207 rue Claude Daunesse, 06904 Sophia-Antipolis Cedex, France}

\author{Roney L. Thompson}
\affiliation{COPPE, Department of Mechanical Engineering, Universidade Federal do Rio de Janeiro, Centro de Tecnologia, Ilha do Fund\~ao, 21945-970, Rio de Janeiro, RJ, Brazil}%

\author{Gilmar Mompean}
\affiliation{Universit\'e de Lille, Polytech'Lille, and Unit\'e de M\'ecanique de Lille, UML EA 7512, Cit\'e Scientifique, 59655 Villeneuve d'Ascq, France}%

\date{\today}

\begin{abstract}
Recent experiments indicated that polymers can reduce the turbulent drag beyond the asymptotic limit known as the MDR, leading to a laminar flow. In this Letter, we show through direct numerical simulations that, when the MDR is exceeded, the flow can remain in a laminar-like regime for a very long period without being truly laminar. During this period called pseudo-laminar state, a transient behavior is observed as a consequence of a small rate of polymer energy injected into the flow. Later on, flow instabilities develop across the channel, finally triggering elastoinertial turbulence. 
\end{abstract}

\maketitle

Thanks to the viscoelastic nature of long chain polymers, their addition, in very small concentrations, can lead to a drastic drag reduction (DR) in turbulent flows. In other words, via a combination of viscous \citep{Lumley-69} and elastic \citep{Tabor-86} effects, polymers modify turbulence, reducing its intensity. 

The DR initially increases with the polymer concentration (elasticity level), but eventually saturates and reaches what is known as the Maximum Drag Reduction Asymptote (MDR) \citep{Virk-67}. The MDR has been interpreted as the edge (or the lower branch solution) between laminar and turbulent motions \citep[][]{Procaccia-08, Xi-12a}, an argument supported by very recent experimental and numerical works carried out with Newtonian fluids \citep{Alizard-19, Pereira-19}. Although the edge is intrinsically unstable in purely Newtonian flows, in viscoelastic scenarios, this marginal turbulent manifestation would persist due to instabilities that emerge from a particular interaction between inertial and elastic forces. For this reason, \citet{Samanta-13} called this flow elastoinertial turbulence (EIT) \citep[see also][]{Dubief-13, Sid-18}.  

Interestingly, recent experimental results pointed out that, for an appropriated range of polymer concentration and Reynolds number, the drag can be reduced to a value beyond the one corresponding to the MDR, leading to a laminar mean velocity profile, suggesting that the corresponding flow is laminar \citep{Choueri-18}. In the following, we show through direct numerical simulations that, in the scenario in which the MDR is exceeded, the flow can remain in a laminar-like state for a very long period of time without being truly laminar. During this period, or flow state referred here as the pseudo-laminar one, a transient behavior is observed as a consequence of a very small rate of polymer energy injected into the mean flow. This polymer/flow transfer of energy gives rise to very weak core-like turbulent structures that later on develop across the channel, finally triggering the elastoinertial turbulence. 


\textit{Formulation:} Turbulent plane Couette flows of incompressible dilute polymer solutions are considered. The flow is driven by both the top and the bottom plates, which have the same magnitude of velocity in the streamwise direction ($U_h$), but opposite signs. The streamwise direction is $x_1 = x$, the spanwise direction is $x_2 = y$, and the wall-normal direction is $x_3 = z$. The instantaneous velocity field in the respective directions is $(u_x, u_y, u_z) = (u, v, w)$ and it is solenoidal ($\nabla \cdot \boldsymbol{u} = 0$, where $\boldsymbol{u}$ denotes the velocity vector). Wall scaling is used and is based on zero-shear rate variables with length and time scaled by $\nu_{tot}/u_{\tau}$ and $\nu_{tot}/u_{\tau}^{2}$, where $\nu_{tot} = \nu_N + \nu_{p0}$ is the total (solvent + polymer) zero-shear viscosity, and $u_{\tau}$ is the friction velocity. Using this scaling, the dimensionless momentum equations are
\begin{equation}
\frac{\partial u_i^+}{\partial t^+} + u_j^+ \frac{\partial u_i^+}{\partial x_j^+} = - \frac{\partial p^+}{\partial x_i^+} + {\beta_0} \frac{\partial^2 u_i^+}{\partial {x_j^+}^2} + \frac{\partial \Xi_{ij}^+}{\partial x_j^+} \, .
\label{eq:mom}
\end{equation}

\begin{figure*}
\centering
\includegraphics[angle=0, scale=0.33]{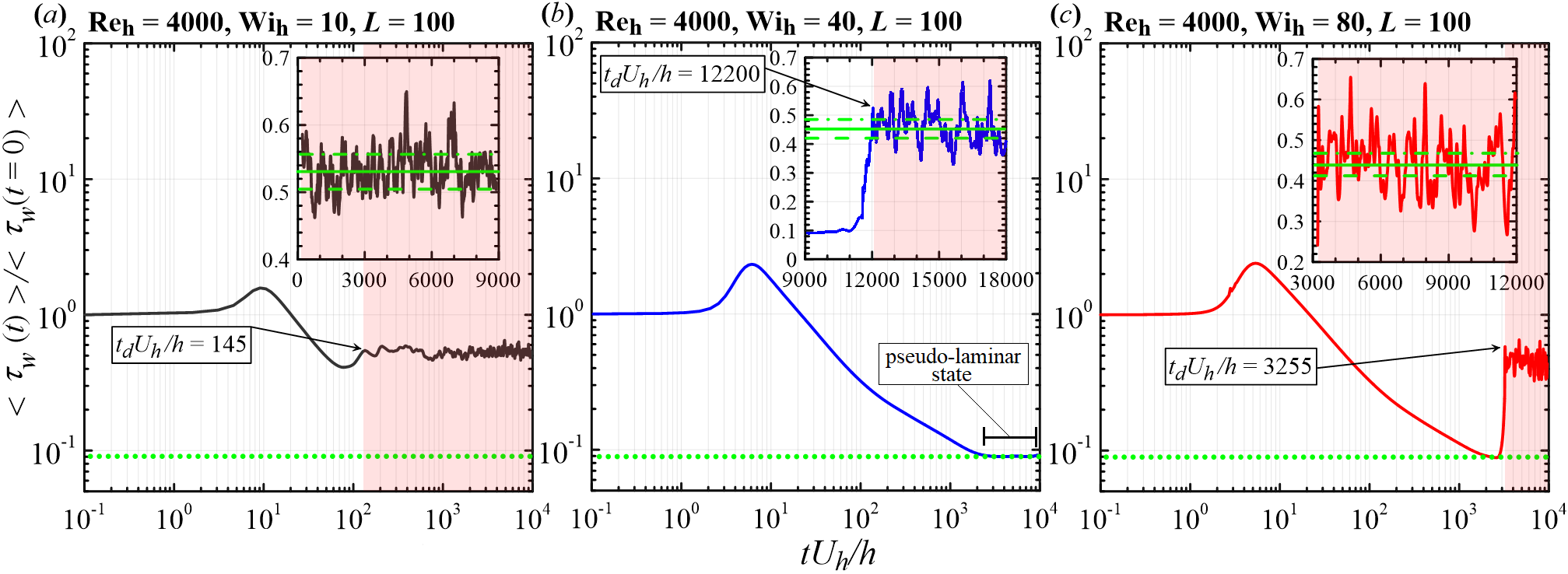}
\vspace*{-0.3cm}
\caption{Evolution in time (made dimensionless by $U_b$ and $h$; $tU_b/h$) of the \textit{x-y} plane average wall shear stress, $\left<\tau_{w}(t)\right>$ (normalized by its initial value value, $\left<\tau_{w}(t = 0)\right>$) for three $\mathrm{Wi_h}$: (\textit{a}) 10; (\textit{b}) 40; and (\textit{c}) 80. The dotted green lines indicate the typical laminar wall shear stress level (made dimensionless by $\left<\tau_{w}(t = 0)\right>$). The insets illustrate the oscillations of $\left<\tau_{w}(t)\right>$ around a statistically steady value represented by the solid black lines. During this final stage, the flow fluctuates between three states (pink regions): hibernating (below the dashed lines); strong activation (above the dash-dotted lines); moderate activation. A movie concerning figure (\textit{a}) is available on url\{https://anselmopereira.net/videos/\}.}
\label{figure-1}
\vspace*{-0.5cm}
\end{figure*} 

In Eq.~\ref{eq:mom}, the superscript `+' indicates the wall unit normalization, $p^+$ is the pressure, $\beta_0$ is the ratio of the Newtonian solvent viscosity ($\nu_N$) to the total zero-shear viscosity ($\nu_{tot}$). The extra-stress tensor components are denoted by $\Xi_{ij}^+$. The formalism of Eq.~\ref{eq:mom} includes the assumption of a uniform polymer concentration in the dilute regime, which is governed by the viscosity ratio $\beta_0$, where $\beta_0 = 1$ yields the limiting behavior of the Newtonian case. The extra-stress tensor components ($\Xi_{ij}^+$) in Eq.~\ref{eq:mom} represent the polymer contribution to the stress of the solution. This contribution is accounted for by a single spring-dumbbell model. We employ here the FENE-P kinetic theory \citet{Bird-87}. The components of the polymeric extra-stress tensor, $\boldsymbol{\Xi^+}$, are then $\Xi_{ij}^+ = \alpha_0 \left( {f \left\lbrace \mathrm{tr}\left(\boldsymbol{C} \right) \right\rbrace C_{ij} - \delta_{ij}} \right)$ with $\alpha_0 = \left(1 - \beta_0 \right) / \mathrm{Wi_{\tau 0}}$, where $\boldsymbol{C}$ denotes the polymeric conformation tensor, and $\mathrm{Wi_{\tau 0}} = \lambda u_{\tau}^2/\nu_{tot}$ is the friction Weissenberg number representing the ratio of the elastic relaxation time ($\lambda$) to the viscous timescale. Additionally, $\delta_{ij}$ is the Kronecker delta and $f\left\lbrace \mathrm{tr}\left(\boldsymbol{C} \right) \right\rbrace$ is given by the Peterlin approximation $f\left\lbrace \mathrm{tr}\left(\boldsymbol{C} \right) \right\rbrace = \frac{L^2 - 3}{L^2 - \mathrm{tr}\left(\boldsymbol{C} \right)}$, where $L$ is the maximum polymer molecule extensibility and $\mathrm{tr}\left(. \right)$ represents the trace operator. This system of equations is closed with an evolution equation for the conformation tensor
\begin{equation}
\begin{split}
\frac{D C_{ij}}{Dt^+} = \left( C_{ik} S_{kj}^{+} + S_{ik}^{+} C_{kj} \right) - \\ 
\left( C_{ik} W_{kj}^{+} + W_{ik}^{+} C_{kj} \right) - \frac{f \left(\mathrm{tr}\left(\boldsymbol{C} \right) \right) C_{ij} - \delta_{ij}}{\mathrm{Wi_{\tau 0}}} \, ,
\end{split}
\label{eq:ct}
\end{equation}
where $S_{ij}^{+} = \left( \partial u_i^+ / \partial x_j^+ + \partial u_j^+ / \partial x_i^+ \right) / 2$ and $W_{ij}^{+} = \left( \partial u_i^+ / \partial x_j^+ - \partial u_j^+ / \partial x_i^+ \right) / 2 $ are, respectively, the components of the rate-of-strain, $\boldsymbol{S^+}$, and the rate-of-rotation, $\boldsymbol{W^+}$, tensors. 

We follow the same numerical method used in \citet{Pereira-17a, Pereira-17b} and all details of the scheme employed are given in \citet{Thais-11}. We simulate the viscoelastic cases fixing the Reynolds number based on the plate velocity, $\mathrm{Re_h} = h U_h/\nu_{tot}$ (where $h$ denotes the plane Couette half-width), at 4000, $\beta_0$ at 0.9 and $L$ at 100. Fourteen viscoelastic cases are studied by setting the following Weissenberg numbers based on the plate velocity ($\mathrm{Wi_h} = \lambda U_h/h$): 2, 4.3, 10, 20, 30, 35, 40, 45, 50, 60, 70, 80, 95, and 110. The respective time-averaged drag reduction values, $DR$, are: 11\% ($\mathrm{Wi_h} = 2$), 33\% ($\mathrm{Wi_h} = 4.3$), 47\% ($\mathrm{Wi_h} = 10$), 52\% ($\mathrm{Wi_h} = 20$), 53\% ($\mathrm{Wi_h} = 30$), 54\% ($\mathrm{Wi_h} = 35$), 55\% ($\mathrm{Wi_h} = 40$) and 56\% ($50 \leq \mathrm{Wi_h} \leq 110$). The Newtonian case at $\mathrm{Re_h} = 4000$ is also considered as a reference ($DR = 0\%$). Lastly, both the size of the domain ($L_x \times L_y \times L_z = 12\pi \times 4\pi \times 2$) and the number of mesh points ($N_x \times N_y \times N_z = 768 \times 512 \times 257$) are kept  fixed for all cases, which leads to a grid resolution of $7.2 \leq \Delta x^+ \leq 9.5$, $3.6 \leq \Delta y^+ \leq 4.8$,  and $0.2 \leq \Delta z^+ \leq 3.4$.  

\textit{Results and discussions:} The initial condition for the conformation tensor is the identity tensor, i.e. $\boldsymbol{C}(t=0) = \boldsymbol{I}$. In addition, for each viscoelastic case, both the velocity and the pressure fields are initiated from the same Newtonian fully developed turbulent flow. We follow then all the polymer/turbulence interaction from the initial Newtonian-like flow (when polymers are coiled) to the final viscoelastic one. As a result of this method, the DR exhibits a marked transient behavior before achieving a more pronounced fluctuating regime with a clear mean value (statistically steady). We define the percentage of DR in time as $DR(t) = \left( 1 - \frac{{<\tau_w}(t)>}{<\tau_w(t = 0)>} \right) \times 100 [\%]$, where $<{\tau_w}(t)>$ is the area-averaged wall shear stress at a given instant $t$ and $<{\tau_w}(t = 0)>$ is the area-averaged wall shear stress at the very beginning of the simulation, when the conformation tensor is isotropic (which corresponds to the case for which polymers are in a coiled configuration). The evolution of $<\tau_w (t)>/<\tau_w(t = 0)>$ as a function of the dimensionless time, $tU_h/h$, is shown in figure \ref{figure-1} for three $\mathrm{Wi_h}$: 10 (\ref{figure-1}\textit{a}); 40 (\ref{figure-1}\textit{b}); and 80 (\ref{figure-1}\textit{c}). Focusing on the $\mathrm{Wi_h} = 10$ case, at the very first instant, the molecules are totally coiled, $tr\left({\boldsymbol{C}} \right)/L^2 \approx 0$, $<\tau_w (t)>/<\tau_w (t = 0)> = 1$, the DR level is then null, and both the turbulent velocity field and structures still exhibit a Newtonian-like nature \citep[see][for the details]{Pereira-17c}. Polymers start to interact with the flow, taking a considerable amount of energy from it, which partially suppresses turbulent structures and increases the wall shear stress that, in turn, achieves its maximum value. Once the flow is weakened, polymers relax, releasing part of the stored energy primarily to the streamwise flow velocity component \citep{Pereira-17d}. As a result, the mean velocity increases, while $<\tau_w (t)>/<\tau_w(t = 0)>$ decreases, reaching its minimal value. Since the mean flow acts as a source of turbulent energy \citep{Thais-13a}, turbulent structures develop, increasing $\tau_w(t)$ and triggering viscoelastic turbulence \citep{Pereira-17c}. The period of time required to trigger the latter is called the developing time, $t_d$ \citep{Pereira-12, Pereira-13, Pereira-17a}. At the final stage, the wall shear stress oscillates around a statistically steady value, $\overline{\left<\tau_{w}\right>}$, represented by the solid green lines (this value was used to compute the time-averaged $DR$ levels presented in the previous paragraph). Since $\mathrm{Wi_h}$ is relatively high (= 10), the flow oscillates between three distinguished states (pink regions): hibernating, Hib (for which $\left<\tau_{w}\right> < 0.95 \overline{\left<\tau_{w}\right>}$; region below the dashed green lines); strong-active, S-Act, which is the counterpart of the hibernation ($\left<\tau_{w}\right> > 1.05 \overline{\left<\tau_{w}\right>}$; region above the dash-dotted green lines); moderate-active, M-Act (for which $0.95 \leq \left<\tau_{w}\right> / \overline{\left<\tau_{w}\right>} \leq 1.05$) \citep{Pereira-19}. The two latter states contain the basic dynamical elements of Newtonian near-wall turbulence, exhibiting a higher drag. In contrast, the turbulent structures almost vanish during the former state, which reduces the drag \citep{Xi-10, Graham-14, Pereira-17c}. 

\begin{figure}
\centering
\includegraphics[angle=0, scale=0.4]{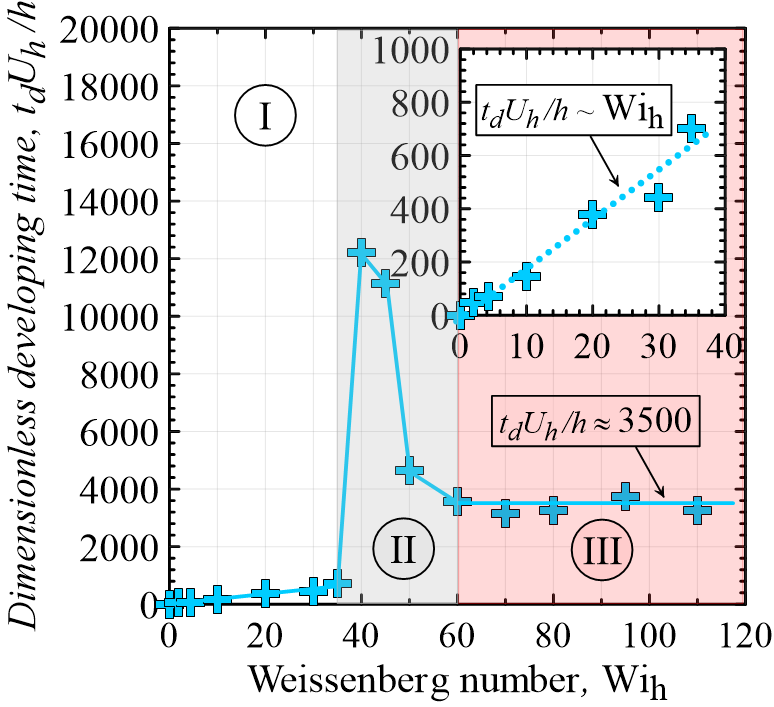}
\vspace*{-0.35cm}
\caption{Dimensionless developing time, $t_d U_h/h$, as a function of $\mathrm{Wi_h}$. Each point represents a simulated case. The inset displays the linear relation $t_d U_h/h \sim \mathrm{Wi_h}$ for $\mathrm{Wi_h} < 40$ (region I, in white). Region II, in gray, indicates the transition towards to EIT, which occurs at $\mathrm{Wi_h} = 40$. For $60 \leq \mathrm{Wi_h} \leq 110$ (region III, in red), $t_d U_h/h$ fluctuates around 3500.}
\label{figure-2}
\vspace*{-0.6cm}
\end{figure} 

Comparing figures \ref{figure-1}(\textit{a}), \ref{figure-1}(\textit{b}), and \ref{figure-1}(\textit{c}), we notice that $t_d U_h/h$ tends to increase drastically from $t_d U_h/h = 145$ to $t_d U_h/h = 12200$ when the Weissenberg number moves from $\mathrm{Wi_h} = 10$ to $\mathrm{Wi_h} = 40$. Conversely, it falls to $t_d U_h/h = 3255$ when $\mathrm{Wi_h} = 80$. The dependence of $t_d U_h/h$ on $Wi_h$ is displayed in details in figure \ref{figure-2}. A fairly linear relation $t_d U_h/h \sim \mathrm{Wi_h}$ (i.e. $t_d \sim \lambda$) is observed for $\mathrm{Wi_h} < 40$ (region I, which appears in white; see the details in the inset). Interestingly, at the critical Weissenberg number of $\mathrm{Wi_h} = 40$, $t_d U_h/h$ increases almost 100 times ($t_d U_h/h = 12200$; region II, in gray), before weakly fluctuating around 3500 for $\mathrm{Wi_h} \geq 60$ (region III, in red). In fact, the abrupt increase of $t_d U_h/h$, from 35 to 40, and its sudden decrease, from 45 to 50, underlined by the gray region in figure \ref{figure-2} indicates the transition towards the EIT, as will be demonstrated in the following lines. 

\begin{figure}
\centering
\includegraphics[angle=0, scale=0.2]{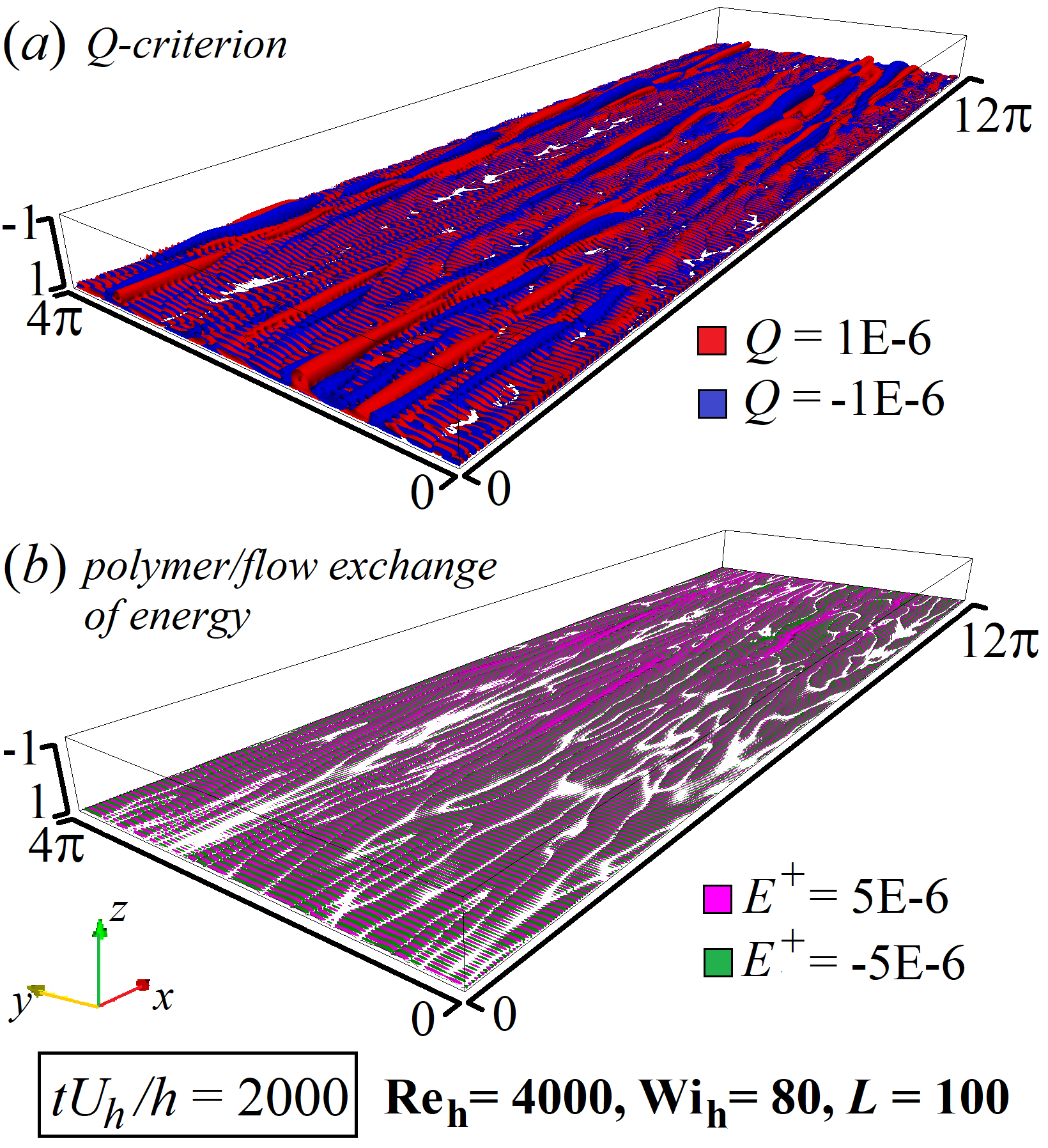}
\vspace*{-0.8cm}
\caption{(\textit{a}) Very weak core-like turbulent structures detected by the $Q$-criterion ($Q = \pm 1\mathrm{E}-6$) at $t U_h/h = 2000$ (pseudo-laminar state). They are organized in trains of alternating elliptical ($Q > 0$; red parts) and hyperbolic ($Q < 0$; blue parts) regions \citep{Dubief-13, Shekar-19} that directly emerge from near-wall polymer/flow exchanges of energy, $E^+$, such as those illustrated in (\textit{b}) ($E^+ = \pm 5\mathrm{E}-6$). A movie showing the development of the core-like turbulent structures is available on url\{https://anselmopereira.net/videos/\}.}
\label{figure-3}
\vspace*{-0.5cm}
\end{figure}   

\begin{figure*}
\centering
\includegraphics[angle=0, scale=0.23]{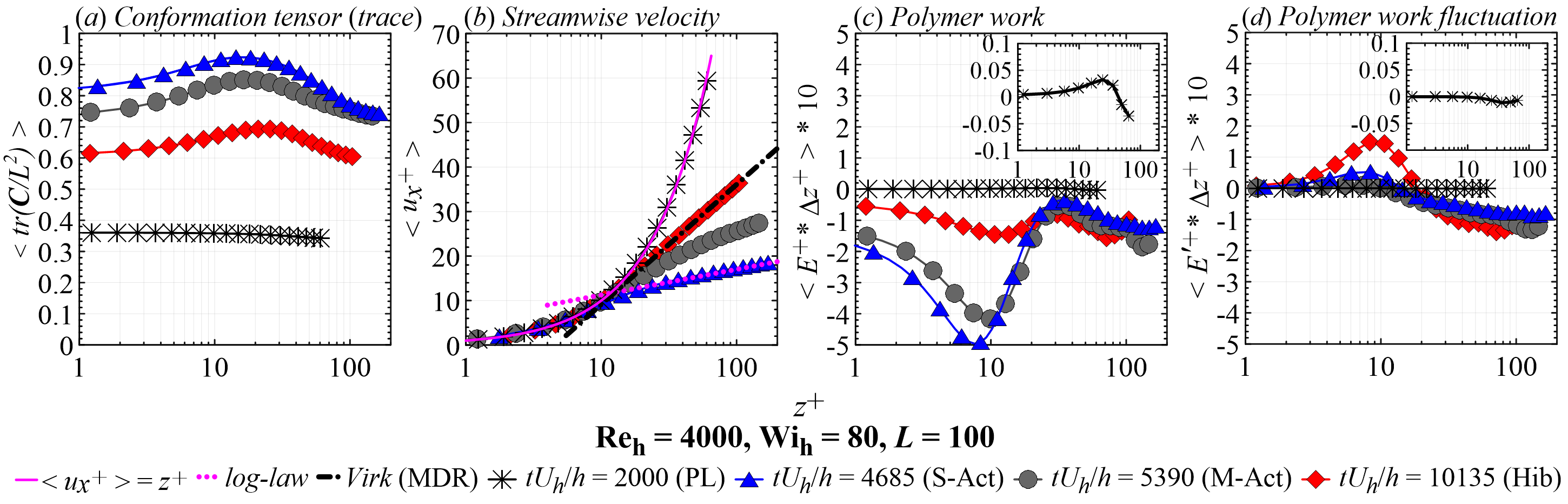}
\vspace*{-0.4cm}
\caption{Normalized $x-y$ plane averaged of the conformation tensor trace (\textit{a}), streamwise velocity (\textit{b}), polymer work (\textit{c}), and polymer work fluctuations (\textit{d}) as a function of normalized wall distance. These profiles are computed at four $t U_h/h$: 2000 (pseudo-laminar), 4685 (strong activation), 5390 (moderate activation), and 10135 (hibernating). For the velocity profile analysis, three references are used: laminar (solid pink line), log-law (dotted pink line), and MDR (dash-dotted black line).}
\label{figure-4}
\vspace*{-0.5cm}
\end{figure*} 

It is important to emphasize that, during the EIT developing process, $\left<\tau_{w} (t)\right>$ decreases considerable, becoming equal to the laminar wall shear stress (denoted by the dotted green lines in figure \ref{figure-1}) before starting to oscillate around its final time-averaged value (indicated by the solid green lines in figure \ref{figure-1}). For the $\mathrm{Wi_h} = 40$ case, for instance, $\left<\tau_{w} (t)\right>$ stays at the laminar level during over 10000 ($2000 < tU_h/h < 12200$). Similarly, the laminar wall shear stress valley is achieved during almost 1000 dimensionless times ($2000 < tU_h/h < 3255$) for the $\mathrm{Wi_h} = 80$ case (figure \ref{figure-1}\textit{c}). At this wall shear stress level, both the normalized conformation tensor trace and the streamwise velocity profiles exhibit a laminar-like dependence on the normalized wall-normal distance, $z^+$, as shown by the black asterisks in figures \ref{figure-4}(\textit{a}) and \ref{figure-4}(\textit{b}), respectively. However, very weak core-like turbulent structures are detected by the Q-criterion of flow classification \citep{Hunt-88}, as shown in figure \ref{figure-3}(\textit{a}) for which $Q = \pm 1\mathrm{E}-6$ and $tU_h/h = 2000$. These are near-wall small scale turbulent structures organized in trains of alternating elliptical ($Q > 0$; red parts) and hyperbolic ($Q < 0$; blue parts) regions which directly emerge from instantaneous near-wall polymer/flow exchanges of energy, such of those illustrated by the pink and green iso-contours corresponding to positive and negative exchange, respectively, in figure \ref{figure-3}(\textit{b}). The polymer/flow energy exchanges are defined as $E^{+} = {u_{x}}^{+} \frac{\partial {\Xi_{x j}}^{+}}{\partial x_j^+} + {u_{y}}^{+} \frac{\partial {\Xi_{y j}}^{+}}{\partial x_j^+} + {u_{z}}^{+} \frac{\partial {\Xi_{z j}}^{+}}{\partial x_j^+} $ \citep[this terms is also called polymer work][]{Dubief-04, Pereira-17b}. Comparing figures \ref{figure-3}(\textit{a}) and \ref{figure-3}(\textit{b}), we clearly observe that the near-wall core-like structures follow the alternating patterns exhibit by $E^{+}$. These particular turbulent structures are only observed here for $\mathrm{Wi_h} \geq 40$. Furthermore, they evolve to the stretched ones aligned with the streamwise direction and located at higher wall distances, leading to the complete development of EIT \citep[see the Appendix of][for more details concerning the final shape of the turbulent structures in non-Newtonian MDR scenarios]{Pereira-19}.                        

We observe then that, at $\mathrm{Re_h} = 4000$, the development of EIT is a process during which the flow can remain in a laminar-like regime for a very long period of time (for instance, $tU_h/h > 10000$ for the $\mathrm{Wi_h} = 40$ case, as shown in figure \ref{figure-2}), without being truly laminar, a particular flow scenario called here pseudo-laminar state (PL). Since this state is characterized by very long developing time scales ($t_d U_h/h > 3000$ for all EIT cases reported in this Letter), an experiment with a duration that is not large enough (typically smaller than $t_d U_h/h$) would not be able to detect the rising of EIT and, consequently, the flow would be classified as laminar due to its signatures. The issue of discrepancy between the time-scales associated with the EIT development and the experiment could be related to the relaminarization argument presented by \citet{Choueri-18}, who investigated experimentally viscoelastic pipe turbulent flows using an opened system for which $tU_D/D < 650$ (where $U_D$ the bulk velocity and $D$ is the pipe diameter) at a bulk Reynolds number of approximately 3000. The findings of the present work suggest that longer experimental investigations would be needed in order to conclude that a laminar regime was in fact achieved. Another possible line of study would be to search for other signatures of the EIT when the laminar profile is achieved, such as hyperbolic and elliptical alternating structures, since a parabolic state would be expected in a laminar regime.

During the pseudo-laminar state, $E^+$ is mainly positive in the very near-wall region (the iso-contours in figure \ref{figure-3}\textit{b} are mostly pink), as shown in figure \ref{figure-4}(c), where the instantaneous $x-y$ plane $\Delta z$-weighted average, $\left< E^+~*~\Delta z^+ \right>$, is displayed as a function of $z^+$ for the $\mathrm{Wi_h} = 80$ case. The $E^+$ curve shown in the inset appears positive for $z^+ < 30$, which indicates that polymers are primarily injecting energy into the very near-wall region at $tU_h/h = 2000$ and, as a result, giving rise to the instabilities that characterize EIT. Despite the laminar signatures of the PL state, very weak flow fluctuations are already detected at $tU_h/h = 2000$, which induces slightly non-null fluctuations of $E^+$ (defined as ${E^{\prime}_{}}^{+} =  {u^{\prime}_{x}}^{+} \frac{\partial {\Xi_{x j}^{\prime}}^{+}}{\partial x_j^+} + {u^{\prime}_{y}}^{+} \frac{\partial {\Xi_{y j}^{\prime}}^{+}}{\partial x_j^+}  + {u^{\prime}_{z}}^{+} \frac{\partial {\Xi_{z j}^{\prime}}^{+}}{\partial x_j^+}$, where the superscript `$\prime$' denotes the fluctuations) at $z^+ > 20$, as indicated by the black asterisks in the inset of figure \ref{figure-4}(\textit{d}). Both the polymer work ($E^+$) and the polymer work fluctuation (${E^{\prime}_{}}^{+}$) profiles change considerably during the final flow stage, when EIT develops and oscillates between S-Act, M-Act and Hib (see the inset in figure \ref{figure-1}\textit{c}). These states are represented by three dimensionless instants in figure \ref{figure-4}: $tU_h/h = 4685$ (S-Act; blue triangles); $tU_h/h = 5390$ (M-Act; gray circles); and $tU_h/h = 10135$ (Hib; red diamonds). During both the S-Act and M-Act states, polymers store a considerable amount of energy from the flow ($\left< {E}^+ \right>$ and $\left< {E^{\prime}}^{+} \right>$ are negative), which increases $\mathrm{tr}\left({\boldsymbol{C}} \right)/L^2$ while $\left< u_x^+ \right>$ decreases towards the log-law profile (dotted pink line in figure \ref{figure-4}\textit{b}) and the velocity fluctuations are dumped (as indicated by the positive values $\left< {E^{\prime}}^{+} \right>$ in both the S-Act and M-Act states). In response, the flow tends to weaken and hibernate. Then, polymers partially relax (red diamonds in figure \ref{figure-4}\textit{a}), releasing energy into the very near-wall flow region, as confirmed by the difference between the red (diamonds) and the blue/gray (triangle/circles) curves in figure \ref{figure-4}(\textit{c}), as well as by the positive values of $\left< {E^{\prime}}^{+} \right>$ illustrated by the red diamonds in figure \ref{figure-4}(\textit{d}). The polymeric injection of energy increases both the mean velocity (red diamonds in figure \ref{figure-4}\textit{b}) and the velocity fluctuations, favoring the re-activation of EIT and the reinitialization of the cycle. Finally, it is important to stress that at the S-Act and Hib states, the velocity profiles are perfectly aligned with the log-law (pink dotted line) and the MDR (black dash-dotted line), respectively, a behavior observed for all EIT cases studied here ($\mathrm{Wi_h} \geq 40$). Such results reinforce the argument according to which the MDR would represent the edge between laminar and non-laminar regimes. 
    
\textit{Conclusion:} In this Letter, we show through direct numerical simulations that, in the scenario in which the MDR is exceeded, the flow can remain in a pseudo-laminar state for a very long period of time. In this state, both the $x-y$ plane averaged velocity and polymer stretching profiles exhibit a laminar feature. Furthermore, the wall shear stress falls to the typical laminar value. However, non-linear polymer/flow interactions give rise to very weak core-like turbulent regions, which slowly amplify and evolve to longer and stretched parts, finally triggering elastoinertial turbulence. The flow then fluctuates between active and hibernating states, while the mean velocity profile oscillates between the log-law (observed during intense activations) and the Maximum Drag Reduction Asymptote (related to pronounced hibernations).   

\textit{Acknowledgements:} This research was granted access to the HPC resources of IDRIS under the allocations 2016-i20162a2277 and i100790 DARI A0032A10304 made by GENCI. Anselmo Pereira would like to thank the PSL Research University for its support under the program `Investissements d'Avenir' launched by the French Government and implemented by the French National Research Agency (ANR) with the reference ANR-10-IDEX-0001-02 PSL.


\bibliographystyle{apsrev4-2}
\bibliography{PRL-ref}

\begin{thebibliography}{25}%
\makeatletter
\providecommand \@ifxundefined [1]{%
 \@ifx{#1\undefined}
}%
\providecommand \@ifnum [1]{%
 \ifnum #1\expandafter \@firstoftwo
 \else \expandafter \@secondoftwo
 \fi
}%
\providecommand \@ifx [1]{%
 \ifx #1\expandafter \@firstoftwo
 \else \expandafter \@secondoftwo
 \fi
}%
\providecommand \natexlab [1]{#1}%
\providecommand \enquote  [1]{``#1''}%
\providecommand \bibnamefont  [1]{#1}%
\providecommand \bibfnamefont [1]{#1}%
\providecommand \citenamefont [1]{#1}%
\providecommand \href@noop [0]{\@secondoftwo}%
\providecommand \href [0]{\begingroup \@sanitize@url \@href}%
\providecommand \@href[1]{\@@startlink{#1}\@@href}%
\providecommand \@@href[1]{\endgroup#1\@@endlink}%
\providecommand \@sanitize@url [0]{\catcode `\\12\catcode `\$12\catcode
  `\&12\catcode `\#12\catcode `\^12\catcode `\_12\catcode `\%12\relax}%
\providecommand \@@startlink[1]{}%
\providecommand \@@endlink[0]{}%
\providecommand \url  [0]{\begingroup\@sanitize@url \@url }%
\providecommand \@url [1]{\endgroup\@href {#1}{\urlprefix }}%
\providecommand \urlprefix  [0]{URL }%
\providecommand \Eprint [0]{\href }%
\providecommand \doibase [0]{https://doi.org/}%
\providecommand \selectlanguage [0]{\@gobble}%
\providecommand \bibinfo  [0]{\@secondoftwo}%
\providecommand \bibfield  [0]{\@secondoftwo}%
\providecommand \translation [1]{[#1]}%
\providecommand \BibitemOpen [0]{}%
\providecommand \bibitemStop [0]{}%
\providecommand \bibitemNoStop [0]{.\EOS\space}%
\providecommand \EOS [0]{\spacefactor3000\relax}%
\providecommand \BibitemShut  [1]{\csname bibitem#1\endcsname}%
\let\auto@bib@innerbib\@empty
\bibitem [{\citenamefont {Lumley}(1969)}]{Lumley-69}%
  \BibitemOpen
  \bibfield  {author} {\bibinfo {author} {\bibfnamefont {J.~L.}\ \bibnamefont
  {Lumley}},\ }\href@noop {} {\bibfield  {journal} {\bibinfo  {journal} {Annual
  Review of Fluid Mechanics}\ }\textbf {\bibinfo {volume} {11}},\ \bibinfo
  {pages} {367} (\bibinfo {year} {1969})}\BibitemShut {NoStop}%
\bibitem [{\citenamefont {Tabor}\ and\ \citenamefont {{de
  Gennes}}(1986)}]{Tabor-86}%
  \BibitemOpen
  \bibfield  {author} {\bibinfo {author} {\bibfnamefont {M.}~\bibnamefont
  {Tabor}}\ and\ \bibinfo {author} {\bibfnamefont {P.~G.}\ \bibnamefont {{de
  Gennes}}},\ }\href@noop {} {\bibfield  {journal} {\bibinfo  {journal}
  {Europhysics Letter}\ }\textbf {\bibinfo {volume} {2}},\ \bibinfo {pages}
  {519} (\bibinfo {year} {1986})}\BibitemShut {NoStop}%
\bibitem [{\citenamefont {Virk}\ \emph {et~al.}(1967)\citenamefont {Virk},
  \citenamefont {Mickley},\ and\ \citenamefont {Smith}}]{Virk-67}%
  \BibitemOpen
  \bibfield  {author} {\bibinfo {author} {\bibfnamefont {P.~S.}\ \bibnamefont
  {Virk}}, \bibinfo {author} {\bibfnamefont {H.~S.}\ \bibnamefont {Mickley}},\
  and\ \bibinfo {author} {\bibfnamefont {K.~A.}\ \bibnamefont {Smith}},\
  }\href@noop {} {\bibfield  {journal} {\bibinfo  {journal} {Journal of Fluid
  Mechanics}\ }\textbf {\bibinfo {volume} {22}},\ \bibinfo {pages} {22}
  (\bibinfo {year} {1967})}\BibitemShut {NoStop}%
\bibitem [{\citenamefont {Procaccia}\ \emph {et~al.}(2008)\citenamefont
  {Procaccia}, \citenamefont {L'vov},\ and\ \citenamefont
  {Benzi}}]{Procaccia-08}%
  \BibitemOpen
  \bibfield  {author} {\bibinfo {author} {\bibfnamefont {I.}~\bibnamefont
  {Procaccia}}, \bibinfo {author} {\bibfnamefont {V.~S.}\ \bibnamefont
  {L'vov}},\ and\ \bibinfo {author} {\bibfnamefont {R.}~\bibnamefont {Benzi}},\
  }\href@noop {} {\bibfield  {journal} {\bibinfo  {journal} {Reviews of Modern
  Physics}\ }\textbf {\bibinfo {volume} {80}},\ \bibinfo {pages} {225}
  (\bibinfo {year} {2008})}\BibitemShut {NoStop}%
\bibitem [{\citenamefont {Xi}\ and\ \citenamefont {Graham}(2012)}]{Xi-12a}%
  \BibitemOpen
  \bibfield  {author} {\bibinfo {author} {\bibfnamefont {L.}~\bibnamefont
  {Xi}}\ and\ \bibinfo {author} {\bibfnamefont {M.~D.}\ \bibnamefont
  {Graham}},\ }\href@noop {} {\bibfield  {journal} {\bibinfo  {journal}
  {Physical Review Letters}\ }\textbf {\bibinfo {volume} {108}},\ \bibinfo
  {pages} {028301} (\bibinfo {year} {2012})}\BibitemShut {NoStop}%
\bibitem [{\citenamefont {Alizard}\ and\ \citenamefont
  {Biau}(2019)}]{Alizard-19}%
  \BibitemOpen
  \bibfield  {author} {\bibinfo {author} {\bibfnamefont {F.}~\bibnamefont
  {Alizard}}\ and\ \bibinfo {author} {\bibfnamefont {D.}~\bibnamefont {Biau}},\
  }\href@noop {} {\bibfield  {journal} {\bibinfo  {journal} {Journal of Fluid
  Mechanics}\ }\textbf {\bibinfo {volume} {864}},\ \bibinfo {pages} {221}
  (\bibinfo {year} {2019})}\BibitemShut {NoStop}%
\bibitem [{\citenamefont {Pereira}\ \emph {et~al.}(2019)\citenamefont
  {Pereira}, \citenamefont {Thompson},\ and\ \citenamefont
  {Mompean}}]{Pereira-19}%
  \BibitemOpen
  \bibfield  {author} {\bibinfo {author} {\bibfnamefont {A.}~\bibnamefont
  {Pereira}}, \bibinfo {author} {\bibfnamefont {R.~L.}\ \bibnamefont
  {Thompson}},\ and\ \bibinfo {author} {\bibfnamefont {G.}~\bibnamefont
  {Mompean}},\ }\href@noop {} {\bibfield  {journal} {\bibinfo  {journal}
  {Journal of Fluid Mechanics}\ }\textbf {\bibinfo {volume} {877}},\ \bibinfo
  {pages} {405} (\bibinfo {year} {2019})}\BibitemShut {NoStop}%
\bibitem [{\citenamefont {Samanta}\ \emph {et~al.}(2013)\citenamefont
  {Samanta}, \citenamefont {Dubief}, \citenamefont {Holzner}, \citenamefont
  {Schafer}, \citenamefont {Morozov}, \citenamefont {Wagner},\ and\
  \citenamefont {Hof}}]{Samanta-13}%
  \BibitemOpen
  \bibfield  {author} {\bibinfo {author} {\bibfnamefont {D.}~\bibnamefont
  {Samanta}}, \bibinfo {author} {\bibfnamefont {Y.}~\bibnamefont {Dubief}},
  \bibinfo {author} {\bibfnamefont {M.}~\bibnamefont {Holzner}}, \bibinfo
  {author} {\bibfnamefont {C.}~\bibnamefont {Schafer}}, \bibinfo {author}
  {\bibfnamefont {A.}~\bibnamefont {Morozov}}, \bibinfo {author} {\bibfnamefont
  {C.}~\bibnamefont {Wagner}},\ and\ \bibinfo {author} {\bibfnamefont
  {B.}~\bibnamefont {Hof}},\ }\href@noop {} {\bibfield  {journal} {\bibinfo
  {journal} {Proceedings of the National Academy of Sciences}\ }\textbf
  {\bibinfo {volume} {110}},\ \bibinfo {pages} {10557} (\bibinfo {year}
  {2013})}\BibitemShut {NoStop}%
\bibitem [{\citenamefont {Dubief}\ \emph {et~al.}(2013)\citenamefont {Dubief},
  \citenamefont {Terrapon},\ and\ \citenamefont {Soria}}]{Dubief-13}%
  \BibitemOpen
  \bibfield  {author} {\bibinfo {author} {\bibfnamefont {Y.}~\bibnamefont
  {Dubief}}, \bibinfo {author} {\bibfnamefont {V.~E.}\ \bibnamefont
  {Terrapon}},\ and\ \bibinfo {author} {\bibfnamefont {J.}~\bibnamefont
  {Soria}},\ }\href@noop {} {\bibfield  {journal} {\bibinfo  {journal} {Physics
  of Fluids}\ }\textbf {\bibinfo {volume} {25}},\ \bibinfo {pages} {110817}
  (\bibinfo {year} {2013})}\BibitemShut {NoStop}%
\bibitem [{\citenamefont {Sid}\ \emph {et~al.}(2018)\citenamefont {Sid},
  \citenamefont {Terrapon},\ and\ \citenamefont {Dubief}}]{Sid-18}%
  \BibitemOpen
  \bibfield  {author} {\bibinfo {author} {\bibfnamefont {S.}~\bibnamefont
  {Sid}}, \bibinfo {author} {\bibfnamefont {V.~E.}\ \bibnamefont {Terrapon}},\
  and\ \bibinfo {author} {\bibfnamefont {Y.}~\bibnamefont {Dubief}},\
  }\href@noop {} {\bibfield  {journal} {\bibinfo  {journal} {Physical Review
  Fluids}\ }\textbf {\bibinfo {volume} {3}},\ \bibinfo {pages} {1} (\bibinfo
  {year} {2018})}\BibitemShut {NoStop}%
\bibitem [{\citenamefont {Choueiri}\ \emph {et~al.}(2018)\citenamefont
  {Choueiri}, \citenamefont {Lopez},\ and\ \citenamefont {Hof}}]{Choueri-18}%
  \BibitemOpen
  \bibfield  {author} {\bibinfo {author} {\bibfnamefont {G.~H.}\ \bibnamefont
  {Choueiri}}, \bibinfo {author} {\bibfnamefont {J.~M.}\ \bibnamefont
  {Lopez}},\ and\ \bibinfo {author} {\bibfnamefont {B.}~\bibnamefont {Hof}},\
  }\href@noop {} {\bibfield  {journal} {\bibinfo  {journal} {Physical Review
  Letters}\ }\textbf {\bibinfo {volume} {120}},\ \bibinfo {pages} {1} (\bibinfo
  {year} {2018})}\BibitemShut {NoStop}%
\bibitem [{\citenamefont {Bird}\ \emph {et~al.}(1987)\citenamefont {Bird},
  \citenamefont {Armstrong},\ and\ \citenamefont {Hassager}}]{Bird-87}%
  \BibitemOpen
  \bibfield  {author} {\bibinfo {author} {\bibfnamefont {R.}~\bibnamefont
  {Bird}}, \bibinfo {author} {\bibfnamefont {R.}~\bibnamefont {Armstrong}},\
  and\ \bibinfo {author} {\bibfnamefont {O.}~\bibnamefont {Hassager}},\
  }\href@noop {} {\emph {\bibinfo {title} {Dynamics of Polymeric Liquids.
  Kinetic Theory}}}\ (\bibinfo  {publisher} {Wiley-Interscience},\ \bibinfo
  {year} {1987})\BibitemShut {NoStop}%
\bibitem [{\citenamefont {Pereira}\ \emph
  {et~al.}(2017{\natexlab{a}})\citenamefont {Pereira}, \citenamefont {Mompean},
  \citenamefont {Thais},\ and\ \citenamefont {Soares}}]{Pereira-17a}%
  \BibitemOpen
  \bibfield  {author} {\bibinfo {author} {\bibfnamefont {A.~S.}\ \bibnamefont
  {Pereira}}, \bibinfo {author} {\bibfnamefont {G.}~\bibnamefont {Mompean}},
  \bibinfo {author} {\bibfnamefont {L.}~\bibnamefont {Thais}},\ and\ \bibinfo
  {author} {\bibfnamefont {E.~J.}\ \bibnamefont {Soares}},\ }\href@noop {}
  {\bibfield  {journal} {\bibinfo  {journal} {Journal of Non-Newtonian Fluid
  Mechanics}\ }\textbf {\bibinfo {volume} {241}},\ \bibinfo {pages} {60}
  (\bibinfo {year} {2017}{\natexlab{a}})}\BibitemShut {NoStop}%
\bibitem [{\citenamefont {Pereira}\ \emph
  {et~al.}(2017{\natexlab{b}})\citenamefont {Pereira}, \citenamefont {Mompean},
  \citenamefont {Thais},\ and\ \citenamefont {Thompson}}]{Pereira-17b}%
  \BibitemOpen
  \bibfield  {author} {\bibinfo {author} {\bibfnamefont {A.~S.}\ \bibnamefont
  {Pereira}}, \bibinfo {author} {\bibfnamefont {G.}~\bibnamefont {Mompean}},
  \bibinfo {author} {\bibfnamefont {L.}~\bibnamefont {Thais}},\ and\ \bibinfo
  {author} {\bibfnamefont {R.~L.}\ \bibnamefont {Thompson}},\ }\href@noop {}
  {\bibfield  {journal} {\bibinfo  {journal} {Journal of Fluid Mechanics}\
  }\textbf {\bibinfo {volume} {824}},\ \bibinfo {pages} {135} (\bibinfo {year}
  {2017}{\natexlab{b}})}\BibitemShut {NoStop}%
\bibitem [{\citenamefont {Thais}\ \emph {et~al.}(2011)\citenamefont {Thais},
  \citenamefont {Tejada-Martinez}, \citenamefont {Gatski},\ and\ \citenamefont
  {Mompean}}]{Thais-11}%
  \BibitemOpen
  \bibfield  {author} {\bibinfo {author} {\bibfnamefont {L.}~\bibnamefont
  {Thais}}, \bibinfo {author} {\bibfnamefont {A.}~\bibnamefont
  {Tejada-Martinez}}, \bibinfo {author} {\bibfnamefont {T.~B.}\ \bibnamefont
  {Gatski}},\ and\ \bibinfo {author} {\bibfnamefont {G.}~\bibnamefont
  {Mompean}},\ }\href@noop {} {\bibfield  {journal} {\bibinfo  {journal}
  {Computers and Fluids}\ }\textbf {\bibinfo {volume} {43}},\ \bibinfo {pages}
  {134} (\bibinfo {year} {2011})}\BibitemShut {NoStop}%
\bibitem [{\citenamefont {Pereira}\ \emph
  {et~al.}(2017{\natexlab{c}})\citenamefont {Pereira}, \citenamefont {Mompean},
  \citenamefont {Thais}, \citenamefont {Soares},\ and\ \citenamefont
  {Thompson}}]{Pereira-17c}%
  \BibitemOpen
  \bibfield  {author} {\bibinfo {author} {\bibfnamefont {A.~S.}\ \bibnamefont
  {Pereira}}, \bibinfo {author} {\bibfnamefont {G.}~\bibnamefont {Mompean}},
  \bibinfo {author} {\bibfnamefont {L.}~\bibnamefont {Thais}}, \bibinfo
  {author} {\bibfnamefont {E.~J.}\ \bibnamefont {Soares}},\ and\ \bibinfo
  {author} {\bibfnamefont {R.~L.}\ \bibnamefont {Thompson}},\ }\href@noop {}
  {\bibfield  {journal} {\bibinfo  {journal} {Physical Review Fluids}\ }\textbf
  {\bibinfo {volume} {2}},\ \bibinfo {pages} {084605} (\bibinfo {year}
  {2017}{\natexlab{c}})}\BibitemShut {NoStop}%
\bibitem [{\citenamefont {Pereira}\ \emph
  {et~al.}(2017{\natexlab{d}})\citenamefont {Pereira}, \citenamefont {Mompean},
  \citenamefont {Thompson},\ and\ \citenamefont {Soares}}]{Pereira-17d}%
  \BibitemOpen
  \bibfield  {author} {\bibinfo {author} {\bibfnamefont {A.~S.}\ \bibnamefont
  {Pereira}}, \bibinfo {author} {\bibfnamefont {G.}~\bibnamefont {Mompean}},
  \bibinfo {author} {\bibfnamefont {R.~L.}\ \bibnamefont {Thompson}},\ and\
  \bibinfo {author} {\bibfnamefont {E.~J.}\ \bibnamefont {Soares}},\
  }\href@noop {} {\bibfield  {journal} {\bibinfo  {journal} {Physics of
  Fluids}\ }\textbf {\bibinfo {volume} {29}} (\bibinfo {year}
  {2017}{\natexlab{d}})}\BibitemShut {NoStop}%
\bibitem [{\citenamefont {Thais}\ \emph {et~al.}(2013)\citenamefont {Thais},
  \citenamefont {Gatski},\ and\ \citenamefont {Mompean}}]{Thais-13a}%
  \BibitemOpen
  \bibfield  {author} {\bibinfo {author} {\bibfnamefont {L.}~\bibnamefont
  {Thais}}, \bibinfo {author} {\bibfnamefont {T.~B.}\ \bibnamefont {Gatski}},\
  and\ \bibinfo {author} {\bibfnamefont {G.}~\bibnamefont {Mompean}},\
  }\href@noop {} {\bibfield  {journal} {\bibinfo  {journal} {International
  Journal of Heat and Fluid Flow}\ }\textbf {\bibinfo {volume} {43}},\ \bibinfo
  {pages} {52} (\bibinfo {year} {2013})}\BibitemShut {NoStop}%
\bibitem [{\citenamefont {Pereira}\ and\ \citenamefont
  {Soares}(2012)}]{Pereira-12}%
  \BibitemOpen
  \bibfield  {author} {\bibinfo {author} {\bibfnamefont {A.~S.}\ \bibnamefont
  {Pereira}}\ and\ \bibinfo {author} {\bibfnamefont {E.~J.}\ \bibnamefont
  {Soares}},\ }\href@noop {} {\bibfield  {journal} {\bibinfo  {journal}
  {Journal of Non-Newtonian Fluid Mechanics}\ }\textbf {\bibinfo {volume}
  {179}},\ \bibinfo {pages} {9} (\bibinfo {year} {2012})}\BibitemShut {NoStop}%
\bibitem [{\citenamefont {Pereira}\ \emph {et~al.}(2013)\citenamefont
  {Pereira}, \citenamefont {Andrade},\ and\ \citenamefont
  {Soares}}]{Pereira-13}%
  \BibitemOpen
  \bibfield  {author} {\bibinfo {author} {\bibfnamefont {A.~S.}\ \bibnamefont
  {Pereira}}, \bibinfo {author} {\bibfnamefont {R.~M.}\ \bibnamefont
  {Andrade}},\ and\ \bibinfo {author} {\bibfnamefont {E.~J.}\ \bibnamefont
  {Soares}},\ }\href@noop {} {\bibfield  {journal} {\bibinfo  {journal}
  {Journal of Non-Newtonian Fluid Mechanics}\ }\textbf {\bibinfo {volume}
  {202}},\ \bibinfo {pages} {72} (\bibinfo {year} {2013})}\BibitemShut
  {NoStop}%
\bibitem [{\citenamefont {Xi}\ and\ \citenamefont {Graham}(2010)}]{Xi-10}%
  \BibitemOpen
  \bibfield  {author} {\bibinfo {author} {\bibfnamefont {L.}~\bibnamefont
  {Xi}}\ and\ \bibinfo {author} {\bibfnamefont {M.~D.}\ \bibnamefont
  {Graham}},\ }\href@noop {} {\bibfield  {journal} {\bibinfo  {journal}
  {Physical Review Letters}\ }\textbf {\bibinfo {volume} {104}},\ \bibinfo
  {pages} {218301} (\bibinfo {year} {2010})}\BibitemShut {NoStop}%
\bibitem [{\citenamefont {Graham}(2014)}]{Graham-14}%
  \BibitemOpen
  \bibfield  {author} {\bibinfo {author} {\bibfnamefont {M.~D.}\ \bibnamefont
  {Graham}},\ }\href@noop {} {\bibfield  {journal} {\bibinfo  {journal}
  {Physics of Fluids}\ }\textbf {\bibinfo {volume} {26}},\ \bibinfo {pages}
  {101301} (\bibinfo {year} {2014})}\BibitemShut {NoStop}%
\bibitem [{\citenamefont {Shekar}\ \emph {et~al.}(2019)\citenamefont {Shekar},
  \citenamefont {McMullen}, \citenamefont {Wang}, \citenamefont {McKeon},\ and\
  \citenamefont {Graham}}]{Shekar-19}%
  \BibitemOpen
  \bibfield  {author} {\bibinfo {author} {\bibfnamefont {A.}~\bibnamefont
  {Shekar}}, \bibinfo {author} {\bibfnamefont {R.~M.}\ \bibnamefont
  {McMullen}}, \bibinfo {author} {\bibfnamefont {S.-N.}\ \bibnamefont {Wang}},
  \bibinfo {author} {\bibfnamefont {B.~J.}\ \bibnamefont {McKeon}},\ and\
  \bibinfo {author} {\bibfnamefont {M.~D.}\ \bibnamefont {Graham}},\
  }\href@noop {} {\bibfield  {journal} {\bibinfo  {journal} {Physical Review
  Letters}\ ,\ \bibinfo {pages} {124503}} (\bibinfo {year} {2019})}\BibitemShut
  {NoStop}%
\bibitem [{\citenamefont {Hunt}\ \emph {et~al.}(1988)\citenamefont {Hunt},
  \citenamefont {Wray},\ and\ \citenamefont {Moin}}]{Hunt-88}%
  \BibitemOpen
  \bibfield  {author} {\bibinfo {author} {\bibfnamefont {J.~C.~R.}\
  \bibnamefont {Hunt}}, \bibinfo {author} {\bibfnamefont {A.~A.}\ \bibnamefont
  {Wray}},\ and\ \bibinfo {author} {\bibfnamefont {P.}~\bibnamefont {Moin}},\
  }\href@noop {} {\bibfield  {journal} {\bibinfo  {journal} {Center for
  Turbulence Research - Proceedings of Summer Program}\ }\textbf {\bibinfo
  {volume} {Report CTR-S88}},\ \bibinfo {pages} {193} (\bibinfo {year}
  {1988})}\BibitemShut {NoStop}%
\bibitem [{\citenamefont {Dubief}\ \emph {et~al.}(2004)\citenamefont {Dubief},
  \citenamefont {White}, \citenamefont {Terrapon}, \citenamefont {Shaqfeh},
  \citenamefont {Moin},\ and\ \citenamefont {Lele}}]{Dubief-04}%
  \BibitemOpen
  \bibfield  {author} {\bibinfo {author} {\bibfnamefont {Y.}~\bibnamefont
  {Dubief}}, \bibinfo {author} {\bibfnamefont {C.~M.}\ \bibnamefont {White}},
  \bibinfo {author} {\bibfnamefont {V.~E.}\ \bibnamefont {Terrapon}}, \bibinfo
  {author} {\bibfnamefont {E.~S.~G.}\ \bibnamefont {Shaqfeh}}, \bibinfo
  {author} {\bibfnamefont {P.}~\bibnamefont {Moin}},\ and\ \bibinfo {author}
  {\bibfnamefont {S.~K.}\ \bibnamefont {Lele}},\ }\href@noop {} {\bibfield
  {journal} {\bibinfo  {journal} {Journal of Fluid Mechanics}\ }\textbf
  {\bibinfo {volume} {514}},\ \bibinfo {pages} {271} (\bibinfo {year}
  {2004})}\BibitemShut {NoStop}%
\end{thebibliography}%

\end{document}